# Critical Doping in Overdoped High-T$_c$ Superconductors

# – a Quantum Critical Point?

J L Tallon[1], J W Loram[2], G V M Williams[1],

J R Cooper[2], I R Fisher[3], J.D. Johnson[2], M.P. Staines[1] and C Bernhard[4]

[1]Industrial Research Ltd, PO Box 31310, Lower Hutt, New Zealand

[2]IRC in Superconductivity, Cambridge University, Cambridge CB3 0HE, United Kingdom

[4]Max Plank Institüt für Festkörperforschung, Stuttgart, Germany

Evidence is presented from the scaling of the Knight shift, entropy and transport properties together with the sharp peaking of condensation energy, critical currents, superfluid density and a variety of other physical properties for the occurrence of a common critical doping point in lightly overdoped high-T$_c$ superconductors (HTS). This critical doping lies at the point where the doping-dependent normal-state pseudogap energy, $E_g$, falls to zero and bears a strong, though incomplete, resemblance to a quantum critical point (QCP). A QCP scenario could lead directly to an explanation of the non-Fermi liquid behaviour of the normal-state metallic phase and the overall generic behaviour of the HTSC.

The high-$T_c$ superconducting (HTS) cuprates are characterised by a generic phase behaviour as a function of doping, evolving from an AF Mott-Hubbard insulator at zero doping, through a low-temperature spin-glass phase to a superconductor with highly unconventional metallic behaviour in the normal state (NS) above $T_c$ and finally to a Fermi liquid at high doping. These features are summarised in Fig. 1. Straddling the spin glass and underdoped superconducting region there exists a wedge-shaped domain stretching high above $T_c$ in which so-called pseudogap correlations suppress spectral weight, strongly reducing the entropy, magnetic susceptibility and, more particularly, all measures of superconductivity including $T_c$. This domain converges on, and pinches off, at about the optimal doping point where $T_c$ maximises revealing the above-noted unconventional metallic behaviour, most notably exemplified by a resistivity which is linear in T to rather high temperatures. Except for the Néel temperature, $T_N$, delimiting the 3D AF state, this phase behaviour for all HTS appears to follow a common function of the doped hole concentration, p, and is representative of the quasi-2D electronic state of the $CuO_2$ planes, whether these planes appear singly as in $La_{2-x}Sr_xCuO_4$ or multiply, per unit cell, as in many other HTSC. Though many theories have been advanced to explain this generic phase behaviour and, in particular, the origin of high-$T_c$ superconductivity, a common consensus is yet to be achieved. Here we focus on the pseudogap (PG) and its intimate connection with superconductivity (SC) showing that the PG energy scale falls almost linearly with p to zero at a common unique point at p = 0.19 in the *lightly-overdoped SC region,* not, we stress, at optimal doping where p = 0.16. We thus distinguish between critical doping (p = 0.19) where superconductivity is most robust and optimal doping (p = 0.16) where $T_c$ maximises. (In the experiments described p is determined either from the thermoelectric power [1] or from the parabolic dependence of $T_c$ on doping [2]). This critical doping state is reflected in sharp maxima in a variety of T = 0 properties, including condensation energy, critical currents and superfluid density, as well as in the scaling of the NS susceptibility $\chi_s$, entropy, S and resistivity. Such behaviour indicates this point could be a *quantum critical point* (QCP). Within the QCP scenario, according to the arguments of Castellani [3] the associated critical fluctuations could be responsible for the strong SC pairing and the unconventional NS behaviour and the overall generic phase behaviour. The resemblance to a QCP is, however, incomplete lacking, in particular, any clear signature of thermodynamic critical behaviour.

The temperature-dependent NS Knight shift $K_s(T)$ typically is depressed at low T in underdoped HTS cuprates (due to the presence of the pseudogap), T-independent in lightly overdoped samples and has a weak, approximately 1/T dependence in overdoped samples.
Fig. 2 shows $^{89}K_s(T)$ plotted as a function of reduced temperature $T/E_g$ for a series of samples
of $Y_{0.8}Ca_{0.2}Ba_2Cu_3O_{7-\delta}$ with different $\delta$ values. As $E_g$ has no meaning for the most overdoped
samples (open symbols) T for each of these is just divided by 195K, the value of $E_g$ at
optimal doping. The scaled data bifurcates into two separate curves for the underdoped and
overdoped regions but the separatrix is not optimal doping at p=0.16 but lies in the lightly
overdoped region at p=0.19. Identical results are obtained for $Y_{0.9}Ca_{0.1}Ba_2Cu_3O_{7-\delta}$. What is
remarkable about this result is the fact that the critical oxygen concentration for the separatrix
differs for the two samples ($\delta$=0.11 for 0.1Ca and $\delta$=0.20 for 0.2Ca) but the doping state,
p≈0.19, is the same. From the scaling of the underdoped data the values of the pseudogap
energy $E_g$ may be determined and these are plotted as a function of p in the insert to Fig. 2
(circles: x=0.1 and triangles: x=0.2). $E_g$ (p) falls more or less linearly to zero at p=0.19.

 Comparable results are obtained from differential heat capacity measurements [4]. A
similar scaling analysis may be carried out for S/T where S is the electronic entropy [5].
Values of $E_g$ thus obtained for $Y_{0.8}Ca_{0.2}Ba_2Cu_3O_{7-\delta}$ are shown by the crosses in the inset to
Fig. 2 and the solid squares in Fig. 3. These can be seen to be almost identical to the values
obtained from the NMR analysis. More generally, $E_g$ values determined in this way for $Y_{1-x}Ca_xBa_2Cu_3O_{7-\delta}$ with x=0, 0.1 and 0.2, $Bi_2Sr_2CaCu_2O_{8+\delta}$ and $La_{2-x}Sr_xCuO_4$ as well as Zn-
substituted samples all fall to zero at p=0.19 [5]. $T_c(p)$ values for $Y_{0.8}Ca_{0.2}Ba_2Cu_3O_{7-\delta}$ are
shown in Fig. 3 by the triangles. We note that the behaviour shown is strongly reminiscent of
quantum critical behaviour seen in a number of heavy fermion systems [6] in which a phase
boundary $T_M$ between two magnetic states falls to zero as a function of pressure. The point
where $T_M \rightarrow 0$ is a QCP and often a "bubble" of superconductivity appears about this point
due to nearly singular interactions which cause the pairing potential to diverge.

 While the nature of the pseudogap is unknown we note that the incommensurate
peaks observed in inelastic neutron scattering near $\mathbf{Q}=(\pi,\pi)$ appear to be intimately related.
For p>0.12 the q-spacing of the satellite peaks is about 0.125 r.l.u. and remains independent
of doping [7]. However, the amplitude of the incommensurate modulation decreases sharply
with increasing doping [7]. We plot in the inset to Fig.2 (solid squares) the p-dependence of
the relative amplitude of the modulation in magnetic scattering intensity for $YBa_2Cu_3O_{7-\delta}$ [7].
This, like $E_g$, falls to zero at p≈0.19. Because these incommensurate peaks have been
attributed to the presence of dynamic stripes of period 8a (consistent with the q-splitting) [8]
this result could indicate a direct connection between the pseudogap and these stripes.

The above-noted heat capacity measurements carried out on the same samples of $Y_{1-x}Ca_xBa_2Cu_3O_{7-\delta}$ show that as $\delta$ is increased (p decreased) the jump, $\Delta\gamma$, in $\gamma$ at $T_c$ remains essentially constant across the overdoped region. Then, starting at $\delta=0.11$ for 0.1Ca and $\delta=0.20$ for 0.2Ca, $\Delta\gamma$ starts to decrease sharply with the opening of the pseudogap, in both cases at p=0.19. The electronic entropy, S(T) may be determined by integrating $\gamma$(T) and the condensation energy is calculated from $U_o = {}^{T_c}(S_{SC} - S_{NS})dT$ where $S_{SC}$ is the entropy in the superconducting state and $S_{NS}$ is the entropy in the normal state. The latter may be determined either by modelling the NS entropy [9] or, quite generally, by scaling S(T) in the normal state as a function of $E_g$ in the same way as noted above for the Knight shift and susceptibility [5]. This allows much of the T-dependence of S in the normal state, including that extrapolated well below $T_c$ and close to T=0, to be fully determined [10]. Fig. 3 shows (diamonds) the doping dependence of the so-determined condensation energy, $U_o$. This is seen to pass through a sharp maximum coinciding with the point p=0.19 where the pseudogap energy falls to zero. Similar results have been found for $Bi_2Sr_2CaCu_2O_{8+\delta}$ [11].

It should be stated that while a key focus in the early development of HTS materials was the magnitude of $T_c$ this is not a good measure of the usefulness of a superconductor. The condensation energy and superfluid density, $\rho_s$, are the key parameters in determining the "strength" of the superconductivity. $\rho_s$, which determines the phase stiffness, governs the rigidity of the SC wavefunction while $U_o$ governs the pinning potential and hence the intragranular critical current density. It is notable that while $T_c$ suffers a small (7%) reduction in progressing from optimal doping to critical doping $U_o$ rises by a further 65%. Direct measurement of the superfluid density using transverse-field muon spin relaxation ($\mu$SR) confirms the same picture [12]. Fig. 4 shows the $\mu$SR depolarisation rate $\sigma$(0) at T=0 plotted as a function of p for samples of $Y_{0.8}Ca_{0.2}Ba_2Cu_3O_{7-\delta}$ with 0, 2 and 4 % Zn substitution for Cu on the $CuO_2$ planes. In these measurements $\sigma$ [$\mu$sec$^{-1}$] = $7.086 \times 10^4 \lambda_{ab}^{-2}$ [nm$^{-2}$] $\propto \rho_s$. Here $\lambda_{ab}$ is the in-plane London penetration depth and, in the following, $\lambda_c$ is the c-axis penetration depth. The data in Fig. 4 show that, for all Zn substitution levels, the superfluid density at T=0 passes through maxima, in each case at p=0.19, and significantly higher than their respective values at optimal doping (p=0.16). In the most heavily overdoped samples the CuO chains are nearly fully oxygenated. Consequently, the superfluid density will be enhanced due to condensation of pairs on the chains [12]. The superfluid density due to the $CuO_2$ planes alone will thus fall more rapidly on the overdoped side than is shown in Fig. 4.

Many models of the irreversibility field have $H_{irr} \propto \lambda_{ab}^{-2} \times \lambda_c^{-2}$ [13] and as a consequence the irreversibility field will pass through a maximum near p=0.19. In fact because $\lambda_c$ increases monotonically with increasing doping the maximum in the irreversibility field is expected a little beyond p=0.19. These expectations will be confirmed below.

A sequence of polycrystalline samples of $Y_{0.8}Ca_{0.2}Ba_2Cu_3O_{7-\delta}$ quenched to different values of $\delta$ were milled to fine powders, graded by sedimentation in isopropyl alcohol then magnetically aligned in epoxy in a field of 11.74 tesla. Magnetisation measurements were carried out in a VSM with the field aligned along the c-axis of the particles. The magnetisation $J_c$ in 0.2 tesla at 10K and 20K is shown in Fig. 5. This is seen to pass through a sharp maximum at p=0.19 as predicted by the p-dependent condensation energy ($J_c$ should vary as $U_o\xi_{ab}$ ). Of course, other extrinsic parameters are also very important in determining the pinning properties of HTS materials. However, for the present series of samples, which are merely progressively deoxygenated, it is unlikely that these have been significantly altered. The irreversibility temperature, $T_{irr}$, at 5 tesla is plotted in Fig. 5 by the diamonds. As anticipated, this also passes through a maximum just beyond p≈0.19.

There are two methods available to suppress superconductivity in order to (i) explore the low-temperature NS properties and (ii) determine the region where superconductivity is most robust. These are impurity scattering by e.g. Zn or Co substitution on the planar Cu sites and the use of intense magnetic fields. We have previously shown that the suppression of $T_c$ by Zn substitution is quantitatively consistent with unitary scattering with a d-wave order parameter across both the underdoped and overdoped regions using, as the only input, values of $\gamma_{Tc}$ determined from heat capacity measurements [14]. Fig. 6 shows the roughly parabolic phase curves $T_c(p)$ for 0, 1, 2, 4, 6, 8 and 10% Co substitution in $Bi_2Sr_2CaCu_2O_{8+\delta}$ [15]. The curves do not collapse symmetrically about optimal doping but along the curve $E_g(p)$ towards the point p=0.19. Similar results are obtained for Zn substitution in $La_{2-x}Sr_xCuO_4$ and $Y_{0.8}Ca_{0.2}Ba_2Cu_3O_{7-\delta}$. In each case the point p=0.19 is the last point of superconductivity for critical substitution in each of the three materials, again confirming that this point, rather than optimal doping, is where superconductivity is most robust.

Boebinger et al [16] have used intense pulsed magnetic fields to suppress superconductivity and determine the NS resistivity to low temperatures. They found a T=0 insulator-to-metal transition occurring at p=0.18-0.19, presumably the same critical doping point that we have noted above (allowing a small level of uncertainty in determining hole concentrations, p). The observed low-temperature logarithmic divergence of the resistivity in the insulating domain p<0.18 has been taken as evidence for one-dimensional transport along stripes but is also consistent with one-dimensional transport in k-space due to the nodal NS gap [17]. As shown in Fig. 7 this family of resistivity data may also be scaled as a function of $T/T_{min}$ into two separate curves, each metallic at high T but one passing through a minimum at $T=T_{min}$ then diverging logarithmically at low T while the other remaining metallic to T=0. The separatrix, again, occurs at p=0.18-0.19. The p-dependence of $T_{min}$ is plotted in the inset to Fig. 7 (circles) together with values (squares) from scaling the NS data of Takagi *et al.* [18]. Evidently $T_{min}$ falls to zero at p=0.19.

The higher-T resistivity, $\rho(T)$, for HTS cuprates displays a generic variation with hole concentration that exhibits a characteristic downward kink at a temperature T* in the underdoped region, a linear-T behaviour around optimal doping and a superlinear behaviour in the overdoped region. Due to thermal expansion effects the precise temperature variation of $\rho(T)$ at very high T has probably not yet been properly determined because the quantity of interest is the resistivity at constant volume. Resistivity data has been measured for a high-quality thin film sample of $Y_{0.7}Ca_{0.3}Ba_2Cu_3O_{7-\delta}$ for a range of values of $\delta$ that span the underdoped and overdoped regions[19]. Following Itoh et al [20] we plot in Fig. 8 the T-dependence of $[\rho(T)-\rho_o]/\alpha T$ for this film where $\rho_o$ is the T=0 intercept of the extrapolated high-T linear region. The high level of Ca in this film means that for $\delta \rightarrow 1$ the film can only be lightly underdoped. Consequently we have included resistivity data for a more underdoped film of pure $YBa_2Cu_3O_{7-\delta}$ (two dotted curves) [21]. Itoh showed that for underdoped samples such plots closely mimic the variation with doping of the Knight shift and susceptibility. The results in Fig. 8 show that this resemblance persists into the overdoped region as well. The downturn of the underdoped data is due to the NS pseudogap and we note that like the susceptibility and S/T, the data shows a low-T saturation for p<0.11 [22]. Notably, for the critically-doped film with p=0.186 (heavy curve) $[\rho(T)-\rho_o]/\alpha T$ remains flat to just 10K above $T_c$ while, for the optimally-doped sample (dashed curve), the downturn due to the pseudogap persists to 175K or more. The small downturn for the critically-doped sample could be due to a small residual pseudogap or more likely to superconducting fluctuations. Clearly such a family of resistivity curves can again be scaled to two separate curves with the separatrix lying at critical doping. A corollary is that the only doping state for which $\rho(T)$ remains linear at all temperatures is critical doping with p=0.19. This is a distinctive feature of a QCP scenario.

The T-Λ phase diagram about a QCP is shown qualitatively in Fig. 9 where Λ is some intensive variable. At low T there lies an ordered phase to the left of the QCP and a quantum disordered phase to the right [3,23]. Above the QCP is the quantum-critical regime in which all energy scales other than temperature are expelled from the problem. In two-dimensions the resistivity is linear in T. Thus at the QCP linearity is preserved to the lowest temperatures as noted for the critically doped sample. The low-T curvature in $[\rho(T)-\rho_o]/\alpha T$ for p<0.19 or p>0.19 would, in this scenario, be associated with the ordered (pseudogap) state and quantum disordered state, respectively. The crossover to linearity at T*(p) and T*′(p) would correspond to the boundaries between the ordered and quantum-critical state, and the quantum-critical and quantum-disordered states, respectively. Precisely this behaviour is suggested by the resistivity data of Fig. 8. The data would be more convincing if it could be extended to higher temperatures, however, oxygen evolution and disordering on the chains preclude this in the $Y_{0.7}Ca_{0.3}Ba_2Cu_3O_{7-\delta}$ samples. For high temperature resistivity data we must turn to $La_{2-x}Sr_xCuO_4$ because of its oxygen stability. Even here, however, the resistivity is measured at constant pressure ( $\rho_p(T)$ ) and must be converted to resistivity at constant volume, $\rho_v(T)$. This has been carried out by Sundqvist and Nilsson [24] using the $\rho_p(T)$ data of Takagi et al. [25] and we reproduce their results in Fig. 10. At medium to low temperature $\rho_v(T)$ shows negative curvature below critical doping and positive curvature above critical doping. In all cases $\rho_v(T)$ is linear at high temperature and at critical doping (p=x=0.20) this linearity extends down to T=0. The p-dependence of the temperature, T*, at which the crossover to linearity occurs is plotted in the inset and T*(p) may be seen to delineate a "phase diagram" very much consistent with the QCP scenario.

Additional evidence for the proximity of a QCP may be found in inelastic neutron scattering studies on optimally-doped $La_{2-x}Sr_xCuO_4$ near $\mathbf{Q}=(\pi,\pi)$ which reveal a magnetic fluctuation amplitude which diverges as $T^{-2}$ with a correlation length that diverges at low-T as $T^{-1}$ [26]. This implies a total correlation weight that diverges as $T^{-1}$. These singular correlations could arise from proximity to the 1/8[th] point or possibly to the p=0.19 critical doping point.

Another feature which appears to be associated with critical doping is the establishment of a full Fermi surface with coherent states persisting out to the zone boundary. In optimal and underdoped HTS cuprates quasiparticle states have short lifetimes near the zone boundary due to strong scattering from AF fluctuations. Because the c-axis tunnelling matrix is strongly weighted to this region of the Fermi surface c-axis transport is incoherent in underdoped cuprates. With the establishment of a full Fermi surface in the neighbourhood of critical doping, and beyond, the c-axis resistivity thus becomes metallic [16] and the carriers are observed to develop a strong $O_{2pz}$ character [27]. The associated reduction in anisotropy and development of three-dimensional coherent transport will have a number of important effects. Most notably within the superconducting state the crossover to three-dimensionality will result in strong pairbreaking for a d-wave order-parameter [28-30]. This could be the reason for the strong reduction in superfluid density for p>0.19 [12,29].

As noted, the resemblance to a QCP scenario is incomplete. The ordered phase would evidently be associated with the pseudogap and in particular possibly associated with the dynamic stripe phases deduced from inelastic neutron scattering, NQR and other studies [23]. However the ordering is not coherent, there is no phase transition associated with the establishment of the pseudogap and there is no order parameter. Perhaps the system is close to a QCP at p=0.19 but in some additional parameter space. Studies on $(La,Eu)_{2-x}Sr_xCuO_4$ show that for p<0.19 an AF phase appears where superconductivity would have been seen in the Eu-free sample [31]. Both the magnitude and doping dependence of $T_N(x)$ is the same as that for $T_c(x)$ in the pure compound. The system reverts to superconductivity for p≥0.20. A QCP has been invoked at p=0.125 for the freezing of dynamic stripes into a spin glass [23]. It is possible then that there may be two QCP's present, one at p=0.125 and another at=0.19 thus accounting for the plateau in $T_c(p)$ that is present in both $La_{2-x}Sr_xCuO_4$ and in $YBa_2Cu_3O_{7-\delta}$ [32].

In conclusion we have presented extensive evidence for the existence of a special doping point in the lightly-overdoped region at p=0.19 where superconductivity is most robust. This point is characterised by a 7% reduction in $T_c$ below its optimal value $T_{c,max}$, a room-temperature thermoelectric power S(290) = -2μV/K and a resistivity that remains linear down to $T_c$. There is a strong similarity to a quantum critical point scenario however we see no evidence for thermodynamic critical behaviour.

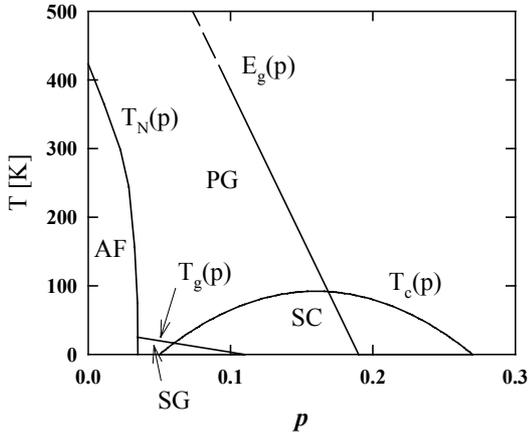

Fig. 1. Schematic phase diagram for the HTS cuprates showing the pseudogap (PG) region and the spin-glass (SG) region. Other terms are defined in the text. The pseudogap energy $E_g$ falls to zero at $p\approx0.19$.

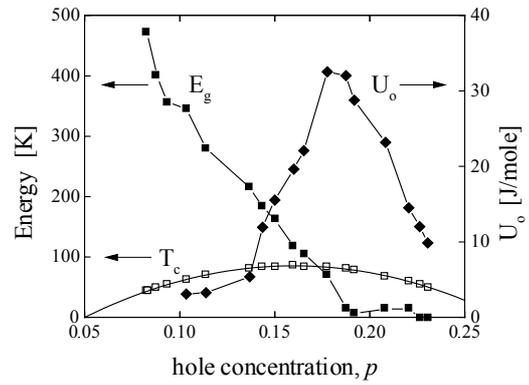

Fig. 3. Values of $E_g$, $T_c$ and condensation energy $U_o$ determined from heat capacity measurements in $Y_{0.8}Ca_{0.2}Ba_2Cu_3O_{7-\delta}$ [4].

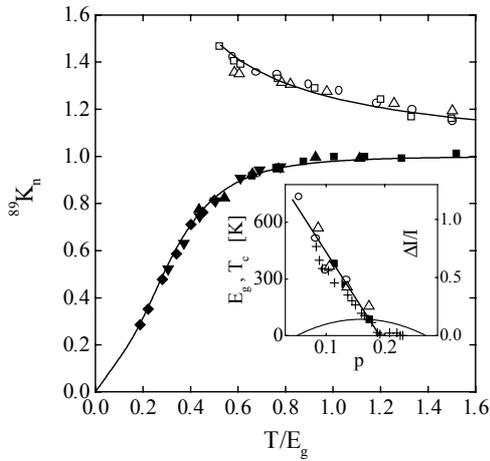

Fig. 2. The $^{89}Y$ Knight shift (divided by 220ppm) for $Y_{1-x}Ca_xBa_2Cu_3O_{7-\delta}$, with x=0.2, plotted as a function of $T/E_g$. Inset: the p-dependence of $E_g$ for x=0.1 (circles) and 0.2 (triangles), and of the relative intensity of the incommensurate INS peaks (squares). $E_g$ values from $C_p$ (+).

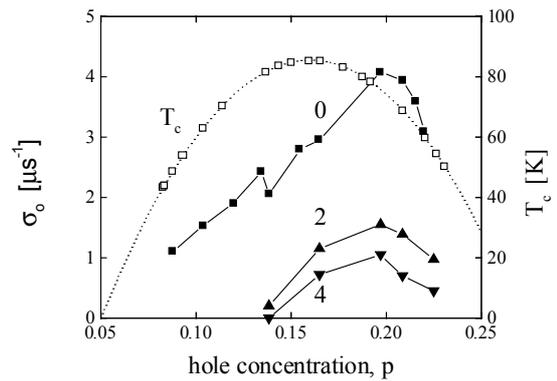

Fig. 4. The p-dependence of the $\mu SR$ depolarisation rate $\sigma_o$ ($\propto$ superfluid density) for $Y_{0.8}Ca_{0.2}Ba_2Cu_3O_{7-\delta}$ with 0, 1, 2, 4 and 6% substitution of Zn on the $CuO_2$ planes.

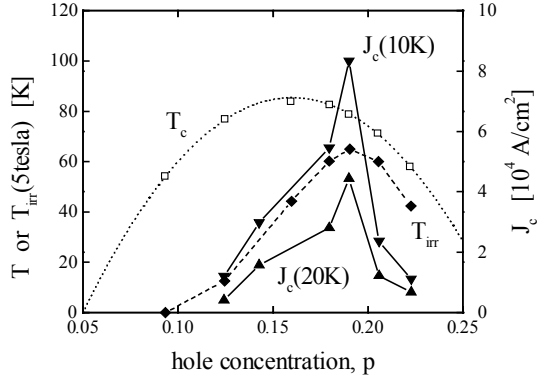

Fig. 5. The p-dependence of $T_c$ (open squares), magnetisation $J_c$ in 0.2 tesla at 10 and 20K (triangles) and irreversibility temperature, $T_{irr}$ at 5 tesla (diamonds) for $Y_{0.8}Ca_{0.2}Ba_2Cu_3O_{7-\delta}$.

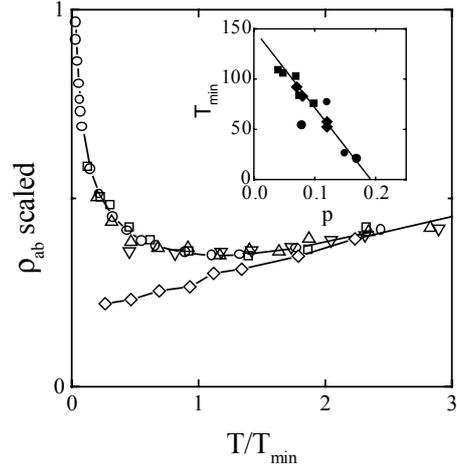

Fig. 7. The resistivity $\rho(T)$ for $La_{2-x}Sr_xCuO_4$ obtained to low T ($<T_c$) by pulsed magnetic fields (O: x=0.08, □: x=0.12, Δ: x=0.15, ∇: x=0.17 and ◇: x=0.22). The data has been scaled and plotted as a function of $T/T_{min}$ where $T_{min}$ is the position of the minimum in $\rho(T)$. The separatrix occurs at p≈0.18. Inset: the p-dependence of $T_{min}$ (●: [16], ■: [18]).

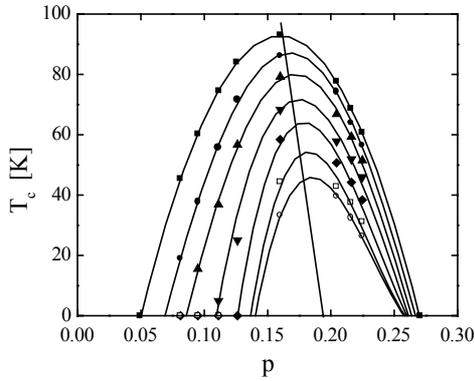

Fig. 6. $T_c(p)$ for a series of Co concentrations (0, 1, 2, 4, 6, 8 and 10 %) in $Bi_2Sr_2CaCu_2O_{8+\delta}$. The sloping line shows the p-dependence of $E_g(p)$.

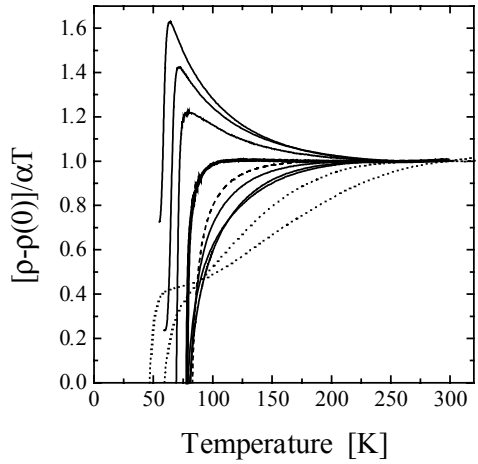

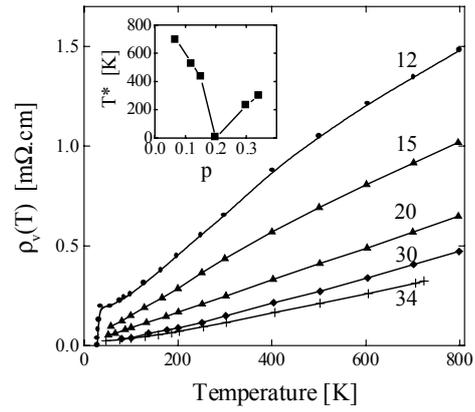

Fig. 8. The resistivity coefficient [ρ(T)-ρ(0)]/αT for $T_c$(p) for thin-film $Y_{0.7}Ca_{0.3}Ba_2Cu_3O_{7-\delta}$ and pure $YBa_2Cu_3O_{7-\delta}$ (dotted curves) with different δ. Heavy curve: critically-doped sample with p=0.186. Dashed curve: nearly optimally doped sample with p≈0.169.

Fig. 10. The T-dependence of the resistivity at constant volume, $\rho_v$(T) for $La_{2-x}Sr_xCuO_4$ (from ref [4]). Inset: the crossover temperature, T*, to high-T linearity in $\rho_v$(T).

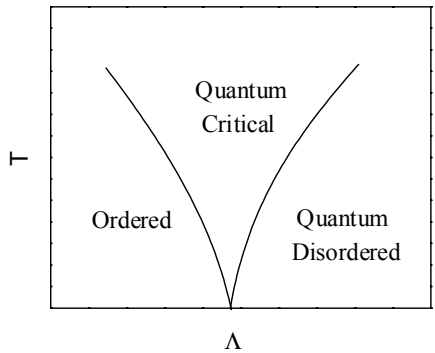

Fig. 9. Schematic T-Λ phase diagram in the neighbourhood of a quantum critical point, QCP.